\documentclass[11pt]{article}
\usepackage{moriond,epsfig}

\def\be{\begin{equation}}
\def\ee{\end{equation}}
\def\bea{\begin{eqnarray}}
\def\eea{\end{eqnarray}}

\begin{document}
%%%%%%%% HAS BEEN SUPPRESSED \vspace*{4cm}
\title{DEPHASING DUE TO NONSTATIONARY $1/f$ NOISE}

\author{\underline{J. SCHRIEFL}$^{1,2}$,
M. CLUSEL$^1$, D. CARPENTIER$^{1}$, P. DEGIOVANNI$^1$, Y. MAKHLIN$^{2,3}$}

\address{$^1$CNRS-Laboratoire de Physique de l'Ecole Normale Sup\'erieure de Lyon, 
69007 Lyon, France}
\address{$^2$Institut f\"ur Theoretische Festk\"orperphysik, Universit\"at Karlsruhe, 76128 Karlsruhe, Germany}
\address{$^3$Landau Institute for Theoretical Physics, Kosygin st. 2, 117940 Moscow, Russia}

\maketitle\abstracts{
Motivated by recent experiments with Josephson qubits we propose a new
phenomenological model for $1/f^\mu$ noise due to
collective excitations of interacting defects in the qubit's environment.
At very low temperatures the effective dynamics of these collective
modes are very slow leading to pronounced non-Gaussian features and 
nonstationarity of the noise.
We analyze the influence of this noise on the dynamics of a qubit in various regimes
and at different 
operation points.
Remarkable predictions are absolute time dependences of a critical coupling and of dephasing 
in the strong coupling regime.
}

Despite a large number of investigations, the detailed explanation of the 
origin of $1/f$ noise stays an open problem in condensed matter physics. 
The diversity of systems that exhibit $1/f$ fluctuations suggests that its 
physical origin is not universal.\cite{Weissman} Still, because of its ubiquity, the study of 
underlying mechanisms, relevant at least in some cases, is interesting.

In the context of quantum-information processing\cite{Chuang}  
the field has attracted new attention.
In particular in solid state devices, e.g. Josephson qubits, low frequency noise, typically with a $1/f$ power spectrum, is considered as the  most important limitation
for the preservation of the phase coherence.  
The microscopic origin of this noise depends thereby on the considered system.   
In the case of charge noise, for instance, it is believed to be
due to the activity of random traps for charges in the
dielectric substrate and in the oxide barrier of the Josephson
junction.\cite{Nakamura2002} 
Similarly, neutral defects in the junction barriers can cause critical current 
fluctuations and low-frequency flux noise may take its origin from trapped 
vortices and magnetic impurities.
In practice, $1/f$ noise appears difficult to
suppress and, since dephasing is dominated by low-frequency noise,
it is particularly destructive.

Decoherence of a qubit due to $1/f$ noise has already been studied 
in
literature using  various models for the noise:
One possibility is to assume Gaussian-distributed fluctuations, e.g. of a large collection
of microscopic modes, which in some cases may be modeled
by a harmonic oscillator bath.\cite{Shnirman2002} 
Alternatively, one can consider a microscopic model with independent bistable fluctuators, each one characterized by a switching rate $\gamma_i$ and an 
effective bias voltage $v_i$ induced at the qubit.\cite{Paladino2002,Galperin} 
If the switching rates are distributed according to $p(\gamma) \sim 1/\gamma$ 
the resulting power spectrum 
exhibits a $1/f$ divergence at low frequencies. 
If only a few fluctuators (microscopic modes) dominate, the effect of the noise differs.

In this work, we present a new model of $1/f$ noise and discuss its effects on 
the phase coherence of a qubit. 
In contrast to the mentioned proposals we consider situations where the mutual
interaction of the localized defects in the background cannot be neglected, 
e.g. at very
low temperatures. 
Since neutral and charged defects behave as elastic and electric dipoles \mbox{respectively 
\cite{Galperin}} the 
strength of the interaction is long-range in both cases decaying as $1/r^3$ as a function of the  
distance $r$.
Although weak, this interaction may play a crucial role at very low temperatures, leading to
formation of highly cooperative clusters of interacting defects.\cite{Burin1995}
The resulting  slow dynamics can be described as a perpetual search of the absolute 
free-energy minima:  the system spends most of the time in local free energy minima corresponding to
metastable configurations, occasionally flipping between these states.
The exponential relation between the typical trapping time $\tau$ and the 
energy barrier to overcome $E$, $\tau(E) \simeq \tau_0 e^{E/T}$, implies that    
 the distribution of the trapping times, $P(\tau)$ is generally broad, decaying very slowly
for large times: $P(\tau) \simeq
\tau^{-1-\mu}$. Here $\mu$ is a positive
dimensionless number that can be written as the ratio of temperature $T$
and a typical height $E_0$ of the energy barriers, $\mu \sim T/E_0$.

The fluctuating extra bias induced at the qubit is described by an asymmetric  
telegraph noise, 
where the  times between two successive plateaux   
are distributed according to an algebraic distribution 
\begin{equation}
P(\tau)= \frac{\mu}{\tau_0} \left(\frac{\tau_0}{\tau_0+\tau}\right)^{1+\mu} \;.
\label{eq:P}
\end{equation}
The heights and lengths of the plateaux are distributed according to narrow 
distributions ${\cal P}(v)$, which is assumed to have zero mean, and ${\rho}(\tau_\uparrow)$ 
respectively. The coupling strength to the qubit is characterized by the 
dimensionless parameter $g^2=\langle (v \tau_\uparrow)^2\rangle$.  
For $\mu > 2$, when $\overline{\tau}=\int_0^\infty d\tau\: \tau P(\tau)$ and 
$\overline{\tau^2}$ are finite our model is equivalent to usual Poissonian telegraph noise as proposed in\cite{Paladino2002,Galperin}.  
However, at ultra-low temperatures, i.e. $\mu<2$, important differences 
emerge.\footnote{Our predictions also differ 
 from those for an ensemble of independent Poissonian fluctuators.\cite{Paladino2002,Galperin}
A detailed comparison is not straightforward and will be addressed elsewhere.}
We thereby have to distinguish two additional classes of noise:
(i) for $1<\mu<2$, $\overline{\tau}$ is still finite but
$\overline{\tau^2}$ diverges. Consequently, the rate of occurring 
plateaux fluctuates strongly around the average $\overline{\tau}^{-1}$. 
In particular, due to the diverging $\overline{\tau^2}$ the statistics of the
noise are only approximately Gaussian, even in the weak coupling regime.
(ii) for $0<\mu<1$, when both  $\overline{\tau}$ 
and $\overline{\tau^2}$ diverge, it is even impossible to define an average rate of 
occurring plateaux, i.e. the noise is scale invariant.  
In this case, the  two point noise correlator decays as
\begin{equation}
  \label{eq:Corr}
  \overline{X(t+t')X(t)} - \overline{X(t+t')}\;\: \overline{X(t)} \quad \simeq\quad 
  \overline{X(t)}^2 \;(t/t')^{1-\mu}\;,
\end{equation}
as opposed to the exponential decay in the case $\mu>2$.
For $0<\mu<1$, the power spectrum  thus exhibits 
a $1/f^\mu$ divergence at low frequencies. 
Furthermore, the $t/t'$-scaling in (\ref{eq:Corr}) implies an intrinsic {\it nonstationarity} of this noise.

In this work we focus on the influence of such a noise on the phase 
coherence of a qubit. 
The full Hamiltonian of a dissipative quantum two-level system can be
written in the form:
\begin{equation}
  \label{eq:Hamiltonian} {\cal H} = -\frac{1}{2}\epsilon \:\sigma_z 
  -\frac{1}{2}\left(\sin\eta \: \sigma_x +  \cos\eta\: \sigma_z\right)X + 
  {\cal H}_{\rm env}(X) \; ,
\end{equation}
where $\epsilon$ and $\eta$ are  control parameters which we assume
to be time-independent and $X$ is the fluctuating extra bias
induced by the environment described by ${\cal H}_{\rm env}$. 
For definiteness we consider in the following 
two special working points for the qubit: 
(i) the optimal point $(\eta = \pi/2)$, 
where the linear longitudinal coupling to the noise is tuned
to zero and consequently decoherence is reduced considerably, \cite{Vion2002} 
and (ii) the case of longitudinal coupling $(\eta =0)$.  
Below we will 
show that transverse linear coupling is 
- regardless of the considered noise model -
equivalent to 
quadratic longitudinal coupling and a remaining transverse noise component with 
suppressed low-frequency tail.
% Within our model it is therefore sufficient to discuss longitudinal fluctuations
%only. 
\vspace*{-0.2cm} 

\paragraph{Optimal point $(\eta = \pi/2)$:} 
In general, transverse noise leads to relaxation, but also contributes to pure dephasing
in higher orders. The effect of low-frequency transverse noise $(\omega \ll \epsilon)$ 
can be treated in the adiabatic approximation: the state of the qubit follows the 
effective magnetic field and (\ref{eq:Hamiltonian}) can be diagonalized to
$-\sigma_z \sqrt{\epsilon^2 + X^2}/2 \approx -\sigma_z \left[\epsilon + X^2/(2\epsilon)\right]/2$. Therefore, low-frequency transverse noise contributes to pure dephasing as 
quadratic longitudinal noise.
On the other hand, higher frequencies, $\omega \sim \epsilon$, mainly contribute
to relaxation and the adiabatic approximation breaks down. 
Situations where both low and high frequencies are present therefore need further 
analysis.

The unitary transformation $U(t) = \exp\{i\Theta(t) \sigma_y/2\}$ with 
$\Theta(t) = \arctan(X(t)/\epsilon) $ transforms the Hamiltonian (\ref{eq:Hamiltonian})
to a spin frame that follows the effective field 
$\epsilon \hat z + X(t)\hat x$:
\begin{equation}
  \label{eq:tildeH}
  \tilde{\cal H} = U(t){\cal H} U^\dagger(t) + i\: U(t) \frac{d}{dt}U^\dagger(t)  = -\frac{1}{2}\sqrt{\epsilon^2+X^2(t)}\; \tau_z + \frac{1}{2} \dot\Theta(t)\: \tau_y  \; ,
\end{equation}
where the Pauli matrices $\tau_i$ remind that the spin is now measured in the 
time-dependent basis, defined by $\tau_z(t) = U(t)\sigma_z U^\dagger(t)$.
Note that for small amplitude of the noise $\left|X(t)\right| \ll \epsilon$ 
the new basis is almost identical to the original basis.  

As we can see from the expression of the transformed Hamiltonian (\ref{eq:tildeH}) 
linear transverse noise is indeed equivalent to quadratic longitudinal noise 
-  as predicted in the adiabatic approximation. 
But due to the explicit time dependence of the transformation $U(t)$ 
an additional transverse component proportional to the time derivative of the 
original noise $X$ appears, 
$\dot\Theta(t) = (\dot X(t)/\epsilon)/(1+(X(t)/\epsilon)^2) \approx \dot X(t)/\epsilon$. 
One might think that the transformed Hamiltonian is therefore even more 
complicated than the original one  since we are now dealing with both longitudinal 
and transverse fluctuations. But a closer look at the expression reveals that 
the low-frequency transverse noise is suppressed,
$S_{\dot\Theta}(\omega) = 2 \langle \dot\Theta(t) \dot\Theta(t')\rangle_\omega \approx \left(\frac{\omega}{\epsilon}\right)^2 S_X(\omega)$. 
Hence, in the experimentally relevant situation where the spectrum $S_X$ is less singular 
than $1/\omega^2$ at low frequencies, $S_{\dot\Theta}$ is a 
regular function at low frequencies with $S_{\dot\Theta}(\omega=0) = 0$. 
As a consequence the transformed transverse noise does not contribute considerably to 
dephasing but only to relaxation.
Dephasing is dominated by the longitudinal part of the noise proportional 
to $X^2$ that still contains all low-frequency contributions. 
We can therefore add up the two contributions in (\ref{eq:tildeH}) independently and
extract simple expressions 
for dephasing and relaxation rates:\vspace*{-0.13cm}
\begin{eqnarray}
  T_1^{-1} & = & \frac{1}{4}S_{\dot\Theta}(\omega=\epsilon) \simeq \frac{1}{4}S_X(\omega=\epsilon)
  \label{eq:T1}\\
  T_2^{-1} & = & \frac{1}{2} T_1^{-1} +  T_\varphi^{-1}
  \label{eq:T2}
\end{eqnarray}
where $T_1$ and the first contribution to $T_2$ are due to the second term in (\ref{eq:tildeH})
and can be calculated simply using the golden rule.
The pure dephasing time
$T_\varphi$ is defined as the characteristic decay time due to quadratic longitudinal noise
$X^2(t)/(2\epsilon)$ in (\ref{eq:tildeH}). In the case of a non-singular noise spectrum it
is given by $T_\varphi^{-1} =  S_{X^2}(\omega=0)/(4\epsilon^2)$. 
For Gaussian $1/f$ noise 
it has been calculated in Ref.\cite{Makhlin2004PRL} In this case the dephasing law differs
from a simple exponential and one should add two decay laws rather than the rates
in (\ref{eq:T2}). 
For the model of nonstationary $1/f$ noise considered in this work the evaluation of $T_\varphi$ 
can be reduced straightforwardly to the case of linear longitudinal coupling:
$X(t)$ and $X^2(t)/(2\epsilon)$ both describe an asymmetric telegraph noise, the major difference being a 
change in the amplitude and thus in the coupling strength.\footnote{In addition $X^2(t)/(2\epsilon)$ has a finite mean. A thorough discussion of this case will be presented elsewhere.}
The coupling strength at the optimal 
point is thereby considerably reduced in comparison to linear longitudinal coupling.\vspace*{-0.15cm}

\paragraph{Longitudinal coupling $(\eta=0)$:}

We now focus on the case of pure dephasing, $\eta=0$ in (\ref{eq:Hamiltonian}).    
Therefore, we follow the time evolution of the off-diagonal entries of the qubit's 
density matrix, 
$\langle \sigma_+(t) \rangle = \overline{D(t_p,t)} e^{i\epsilon t}\langle \sigma_+(0) \rangle$,
where the dephasing factor is defined by  $\overline{D(t_p,t)} = \overline{e^{i\Phi(t_p,t)}}$ 
with $\Phi(t_p,t) = \int_{t_p}^{t_p+t}X(t')dt'$.
Because of the intrinsic nonstationarity of the noise (\ref{eq:Corr}) 
we expect to find an explicit dependence of dephasing on the 
{\it preparation time} $t_p$.
In the following we will therefore investigate this explicit dependence of  
$\overline{D(t_p,t)}$ on $t_p$. 
This can be achieved using CTRW techniques 
and taking special care of initial conditions at $t_p$.\cite{lettre}

As already discussed previously we have to distinguish three different 
cases depending on the value of $\mu$:
When $\mu>2$ the noise is equivalent to Poissonian telegraph noise
\cite{Paladino2002} and
therefore stationary (for $t_p>\tau_0$). 
Consequently, dephasing is insensitive to the absolute time $t_p$.

\begin{figure}[t]
  \centering
  \includegraphics[width=0.29\textwidth,angle=270]{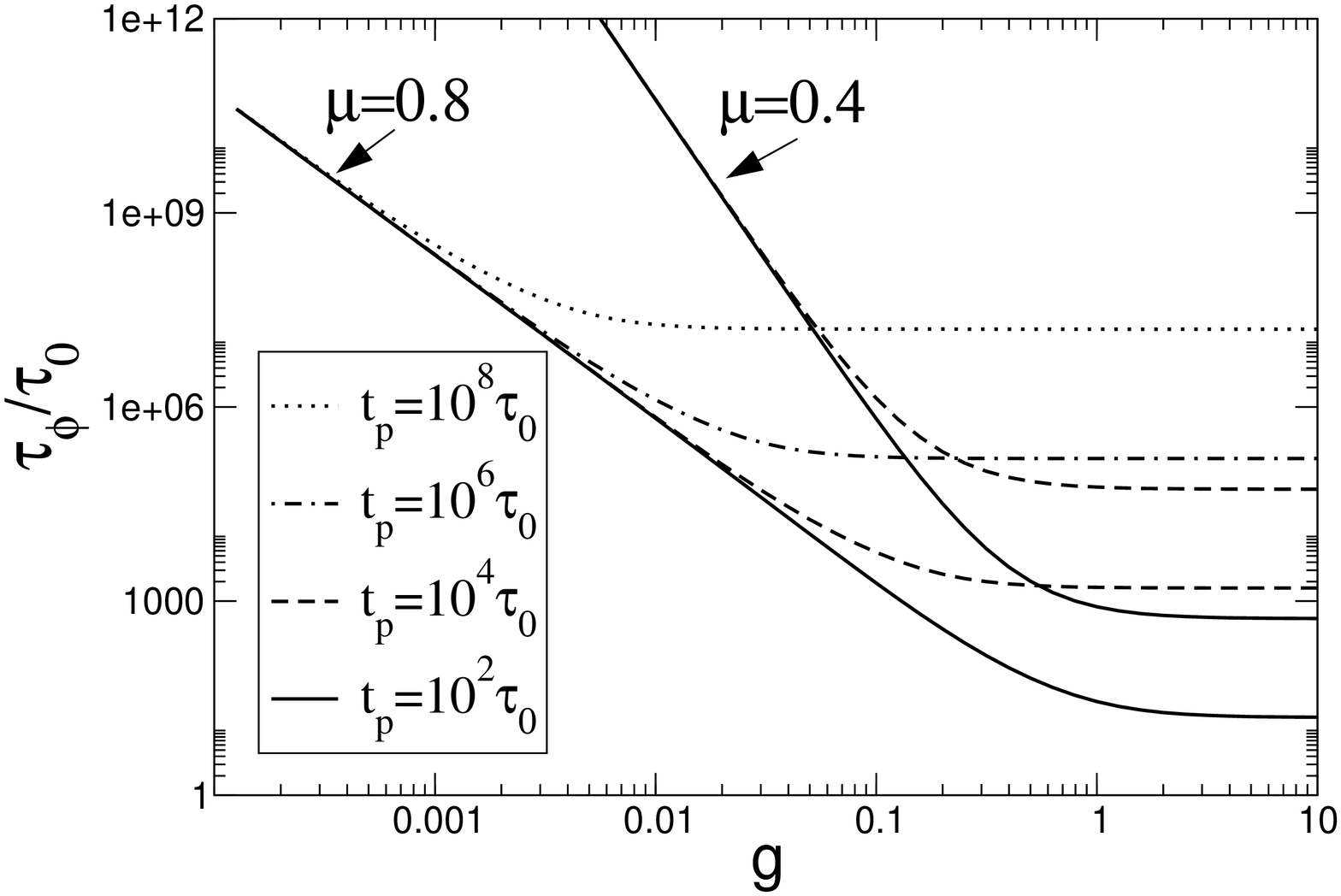}
  \includegraphics[width=0.29\textwidth,angle=270]{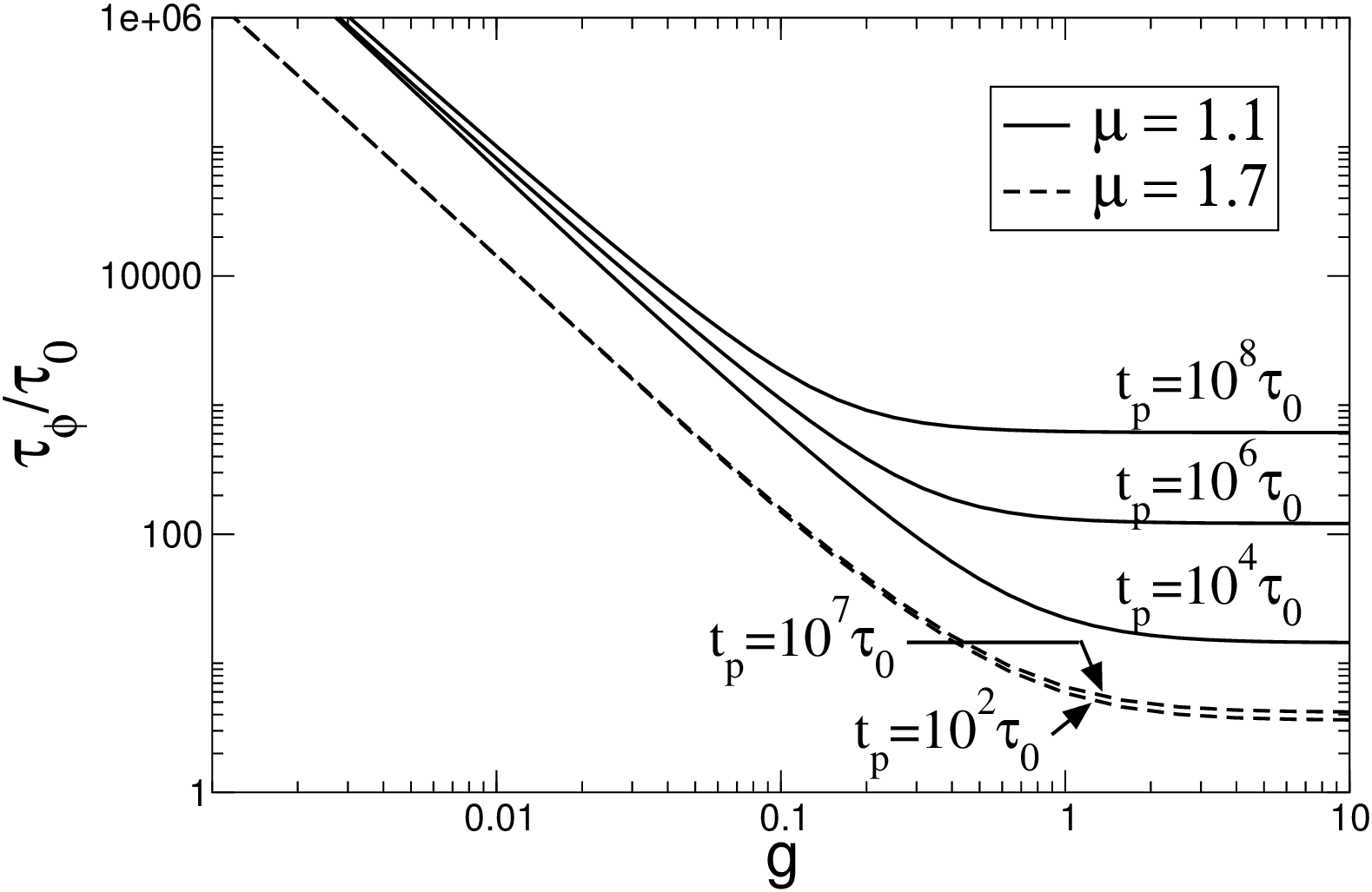}
    \vspace*{-0.4cm}
  \caption{Dephasing time $\tau_\phi$ as a function of the coupling strength $g$ for 
$\mu<1$ (left) and $\mu > 1$ (right).}
    \vspace*{-0.3cm}
  \label{fig}
\end{figure}

For $1<\mu<2$ it turns out that nonstationarity only plays a role in the 
strong coupling regime $g>1$. For weak coupling $g<1$ the noise is 
qualitatively
equivalent to the Poissonian case $\mu>2$, i.e. only subdominant 
corrections appear.\cite{lettre}
 For $g>1$ the dephasing law differs from a simple exponential and the dephasing time, defined as
 $\overline{D(t_p,\tau_\phi)} = e^{-1}$, scales as 
$\tau_\phi \simeq \tau_0 [1/e + (\tau_0/(\tau_0+t_p))^{\mu-1}]^{-1/(\mu-1)}$.
The dependence of $\tau_\phi$ on $t_p$ thus only matters for values of 
$\mu$ close to 1 and disappears as $\mu$ increases to higher values (see Figure \ref{fig}).
   
As expected nonstationarity turns out to be most prominent in the case $0<\mu<1$. 
Again, we have to distinguish two regimes of weak and strong coupling
separated by a critical coupling constant $g_c$. 
But here $g_c$ is not of order one, but depends explicitly on 
the age of the noise: 
$g_c(t_p) \simeq \lambda(\mu) t_p^{-\mu/2} \ll 1$.   
Therefore, the range of the strong coupling regime increases with the age
of the system implying that any qubit subject to a noise with $0<\mu<1$ 
eventually ends up in the strong coupling regime.

For weak coupling $g<g_c(t_p)$ 
dephasing is essentially insensitive to the age of the noise. 
Away from a (short) initial aging regime $\overline{D(t_p,t)}$ decreases as 
$\exp(-(t/\tau_\phi)^\mu)$, where the dephasing time scales as 
$\tau_\phi \sim \tau_0 g^{-2/\mu}$ (in the marginal case $\mu=1$ we get 
logarithmic corrections $\tau_\phi \sim \tau_0 g^{-2}|\ln g|$).  
Indeed, for $t<\tau_\phi$, $\overline{D(t_p,t)}$ is accurately given 
within a Gaussian approximation of the phase: 
$\overline{D(t_p,t)} \simeq \exp(-\frac{1}{2} \overline{\Phi(t_p,t)^2})$.
Only for $t>\tau_\phi$ the Gaussian approximation breaks down and
the exponential decay goes over to a much slower algebraic one, 
$\overline{D(t_p,t)} \propto 1/t^\mu$. 

In the strong coupling regime $g>g_c(t_p)$ the dephasing depends strongly
on the age of the noise. 
The Gaussian approximation 
of the phase 
breaks down even at short times and the decay is not exponential any more.
The dephasing time  becomes proportional to 
$t_p$, $\tau_\phi \simeq C(\mu) t_p$, and saturates as a function of $g$. 
We emphasize that for $\mu \to 1$, 
$C(\mu)$ takes smaller and smaller values implying that the dephasing 
time can be orders of magnitude shorter than the preparation time.    

In summary, we have proposed a new model of $1/f^\mu$ noise describing
phenomenologically a slow collective environment. 
The present model shows a cross-over from 
nonstationary $1/f^\mu$ noise to a memoryless
Poissonian telegraph noise as a function of temperature.
We analyzed the decay
of coherence of a qubit subject to this noise at different operation points.
For low temperatures and strong coupling the dephasing time was found to 
depend explicitly
on the age of the noise.

\end{document}